\documentclass[
,secnumarabic,amssymb, amsmath, nofootinbib,tightenlines,
nobibnotes, aps, prl]{revtex4}
\usepackage{amsmath}%
\usepackage{longtable}%
\usepackage{bm}%
\expandafter\ifx\csname package@font\endcsname\relax\else
 \expandafter\expandafter
 \expandafter\usepackage
 \expandafter\expandafter
 \expandafter{\csname package@font\endcsname}%
\fi

\usepackage{graphicx}
\usepackage{color}

\begin{document}

\title{ Anomalous Exponents  in Strong Turbulence}
 \newcommand {\BU}{Department of Mechanical Engineering, Boston University, Boston, Massachusetts, 02215,USA}

\author{Victor Yakhot}
\affiliation{\BU}
\author{ Diego A.~Donzis }
 \affiliation{Department of Aerospace Engineering, Texas A\&M University, College Station, Texas 77843,USA}
\email[]{donzis@tamu.edu}



\date{\today}

\begin{abstract}

 \noindent   To characterize  fluctuations in a  turbulent flow, one
 usually studies  different  moments of velocity increments  and
 dissipation rate,     $\overline{(v(x+r)-v(x))^{n}}\propto
 r^{\zeta_{n}}$ and $\overline{{\cal E}^{n}}\propto Re^{d_{n}}$,
 respectively.  In high Reynolds number flows,  the moments of different
 orders 
 cannot be simply related to each other which is
 the signature  of  anomalous scaling, one of the most puzzling features
 of  turbulent flows.   High-order moments are related to extreme,  rare
 events and our ability to quantitatively describe  them is crucially
 important for meteorology, heat, mass transfer and other applications.
 In this work we present a solution to this problem in the particular case
 of the Navier-Stokes equations driven by a random force.  A novel
 aspect of this work is that, unlike previous efforts which aimed at
 seeking solutions around the infinite Reynolds number limit, we
 concentrate on  the  vicinity of transitional Reynolds numbers
 $Re^{tr}$ where the first emergence of anomalous scaling is observed 
 out of a low-$Re$ Gaussian background. 
 The obtained closed expressions for anomalous
 scaling exponents $\zeta_{n}$ and $d_{n}$, which depend on 
 the transition Reynolds number,
 agree well with 
 experimental  and numerical data in the literature  and,  when $n\gg 1$,
 $d_{n}\approx  0.19n \ln(n)$.    The theory yields  the energy spectrum
 $E(k)\propto k^{-\zeta_{2}-1}$  with $\zeta_{2}\approx 0.699$,
 different from the outcome of Kolmogorov's  theory.  It is also argued
 that  fluctuations of dissipation rate and those of the transition
 point itself are  responsible for both, deviation from Gaussian
 statistics and multiscaling of velocity field.  

 \end{abstract}

\maketitle
PACS numbers 47.27\\

\noindent {\bf   I. Introduction.} \\
 Thanks to rapid development of experimental and computational methods,  Kolmogorov's dimensional considerations  of 1941,  leading to  the energy spectrum 
$E(k)\approx {\overline{\cal
E}^{\frac{2}{3}}k^{2}\overline{|v(k)|^{2}}}\propto k^{-\frac{5}{3}}$,
have   been  reasonably well  supported by both physical and numerical
experiments.  Still,  one cannot  rule out some slight correction  to
the $5/3$ exponent. In this  formula, the dissipation rate ${\cal
E}=2\nu S_{pq}S_{qp}$  with
$S_{pq}=(\partial_{p}v_{q}+\partial_{q}v_{p})/2$,  is a fluctuating
small-scale parameter and appearance of $\overline{{\cal E}}$ in
Kolmogorov's energy  spectrum can be traced to  Kolmogorov's   relation
$\overline{(v(x+r)-v(x))^{3}}= -\frac{4}{5}\overline{{\cal E}}r$,  where
$v$ is the $x$-component of velocity field and $r$ is the displacement
in the $x$-direction  chosen in the inertial range (IR)  $\eta\ll r\ll
L$. Here  $\eta$ and $L$ denote viscous  and integral length-scales
($L)$, respectively.  The energy balance in  the Navier-Stokes equations
gives $\overline{{\cal E}}={\cal P}$ where ${\cal P}=const$ denotes the
power of external large-scale energy source.
Based on a bold  assumption of both ultra-violet  ($\eta )$ and infra-red ($L$)  cut-offs disappearance in the IR  dynamics  and on his  own analysis of available  experimental data, Kolmogorov  came up with his famous expression for the  energy spectrum.  
It is clear that,  in principle,  this assumption may not be correct and
fluctuations of dissipation rate can somewhat modify the $5/3$ exponent
and, moreover,  lead to dramatic effects in all multi-point and high-order
correlation functions.  Kolmogorov's theory (K41) was subsequently 
generalized 
with various ``cascade models''  leading to
$S_{n}=\overline{(\delta_{r}v)^{n}}\equiv
\overline{(v(x+r)-v(x))^{n}}\propto r^{\zeta_{n}}$ with  $\zeta_{n}=n/3$.
Later, it became clear that   the exponents $\zeta_{n}=n/3$,   contradicted
experimental data pointing to  
the existence of  much more complex relations between $\zeta_{n}$ and
$\zeta_{m}$ with $m\neq n$. 

\noindent In this paper we address the problem of  the scaling exponents  in a simplified setup of an infinite fluid driven by a Gaussian white-in-time random force acting  in the range of scales $r\approx L=O(1)$. When  the amplitude of the stirring force is very small, the flow is governed by   linear  contributions to the Navier-Stokes equations   and the resulting random velocity field  is Gaussian.  
With an increase in the forcing amplitude, at some Reynolds number
($Re^{tr}$), which we loosely call ``transitional'' (see below),   the
nonlinear and linear terms become comparable  and,   eventually,  non-linearity
dominates  and  deviations from Gaussianity take over. This behavior  has been
demonstrated in Refs.~[1,2] and the transition itself was discussed in terms of
breakdown of Random Phase Approximation (RPA)  in Ref.~[3].  

In 1965, in his classic textbook [4],  R.P. Feynman proclaimed  turbulence ``a
central problem,   put aside by   physicists more than a century ({\it
nineteenth}!) ago'',    which one day we will ``have to solve and we do not
know how.'' 
His point was that   by 1965 we knew more about weak and strong interactions
of elementary particles  than about  flow of water out of   faucets  in our
bathtubs.  
Feynman  did not elaborate  on his  statement and we  can only guess that  he had in mind a  derivation of chaotic,  high-Reynolds number   solutions  directly from  deterministic and well-known Navier-Stokes equations for incompressible fluids.  
It  also became clear (for a comprehensive review see Refs.~[5])  that, due to the  lack of a small coupling constant,  the problem of  derivation of the  energy spectra  and other correlation functions from the NS equations using renormalized perturbation expansions was   as hard as  that  in the  theory of strong interactions for which important advances had been achieved [6]. Today, in the 21th century, the problem is still open. \\
  

\noindent In this paper we develop a theory leading to an 
analytic evaluation of
scaling exponents of moments of velocity derivatives including those of
dissipation rate and the exponents $\zeta_{n}$  of the structure functions
$S_{n}(r)=(\overline{(v(x+r)-v(x))^{n}}\propto r^{\zeta_{n}}$.  
An implication of the theory is 
that anomalous scaling is a consequence of dynamic constraints on 
high-order moments of velocity derivatives, following directly from the
behavior of 
Navier-Stokes solutions at low Reynolds numbers.\\

This paper is organized as follows:  in Section II  we introduce an
infinite number of $n$-dependent Reynolds numbers reflecting the multitude
of anomalous exponents characterizing moments of different order $n$.  In
Section III, the Navier-Stokes equations driven by a random Gaussian force are
introduced as a basic model treated in the paper. It is shown  there that in
the low-Reynolds number limit  $Re\ll Re^{tr}$  the solution is Gaussian while as
$Re\gg Re^{tr}$, the moments of  velocity derivative are given by scaling
functions with unknown anomalous exponents and amplitudes. The
crossing of
these two limiting curves at $Re=Re^{tr}$ gives an equations  for
the unknown
anomalous exponents. In Section IV  a transitional Reynolds numbers
$R^{tr}_{\lambda,n,}$  from  weakly turbulent low-Reynolds number
Gaussian
fluid to  the high-$Re$ anomalous state  is discussed. In Section V the exponents
are calculated from the derived equations and compared with experimental and
numerical data. Section VI is devoted to summary and discussion. \\

\noindent {\bf II. Local Reynolds numbers and transitions.
Definitions.}  We are interested in an infinite fluid stirred by a
Gaussian random force acting on a scale  $r\approx L$. Thus, even in the
limit of very weak forcing (for quantitative definitions see below) the
generated flow is  random.  To characterize this class of flows one
typically uses a Reynolds number $Re=(\delta_{r}v
)_{rms}L/\nu\approx v_{rms}L/\nu$. In the ``weak turbulence range'',
when  the forcing amplitude tends to zero and $Re\ll Re^{tr}$,  the PDF
of the generated velocity field  is close to Gaussian. 
With increase of  the Reynolds number to $Re\approx Re^{tr}$,  the flow
undergoes a transformation manifested by  appearance  of a broad,
non-Gaussian,  tails of the probability density  [1-2],  sometimes
associated with breakdown of the Random Phase Approximation (RPA) and a
phase organization  discussed in [3]. 
Since the background flow is random, it is clear that even at
$Re<Re^{tr}$, there exist  low-probability realizations with local
Reynolds number $Re\geq Re^{tr}$ where the flow is turbulent [1].
With further increase of $Re$-number  the Gaussian central part of the
PDF disappears [1,2].  Thus, in general, turbulent flows can be a
superposition of weak and strong turbulence patches.

If  a random  field, for example a velocity gradient,   obeys   Gaussian
statistics with a $Re$-dependent variance 
$\overline{(\partial_{x}v_{x})^{2}}\propto
\frac{v_{0}^{2}}{L^{2}}Re^{\rho_{2}}$, then its moments are given by 
$\overline{(\partial_{x}v_{x})^{2n}}=(2n-1)!!( \overline{(\partial_{x}v_{x})^{2}})^{n}
\propto \frac{v_{o}^{2n}}{L^{2n}}Re^{n\rho_{2}}$.  
Here, $v_{0}=O(1)$ and $L=2\pi/k_{0}=O(1)$ are  the  single-point large-scale properties of the flow. 
The above-mentioned ``normal scaling''  is not the only possibility.
Indeed,  high-Reynolds-number flows are characterized by ``anomalous''
scaling exponents reflecting the formation  of coherent structures. In
this case $\overline{(\partial_{x}v_{x})^{2n}}\propto
\frac{v_{0}^{2n}}{L^{2n}}Re^{\rho_{2n}}$ where $\rho_{2n}\neq
n\rho_{2}$.  

\noindent High-order moments of velocity increments or dissipation rate
characterize rare,  extreme events which make the knowledge of
anomalous scaling exponents very important. In a random flow with
$Re< Re^{tr}$ there always exist some
statistical realizations with {\it local} $Re_{n}>Re^{tr}$ which are
turbulent. This effect has been supported by studies in isotropic and
homogeneous turbulence  [1] and in a ``noisy'' flow in micro-channels
[2].  Thus,  the low-$Re$ number flow can be a superposition of a Gaussian
and anomalous, strongly turbulent, patches [1,2]. To account for
these effects, an infinite number of mean fields
$\hat{v}(n,m)=L\overline{(\partial_{x}v_{x})^{n}}^{\frac{1}{m}}$  are
defined for convinence which lead to corresponding Reynolds numbers:

 \begin{equation}
\hat{R}e_{n}= \hat{v}(n,n)L/\nu\equiv
\frac{L^{2}\overline{(\partial_{x}v_{x})^{n}}^{\frac{1}{n}}}{\nu}
=
A_n Re^{\frac{\rho_{n}}{n}+1}
\end{equation}

\noindent where $\rho_{n}$ is a scaling  exponent of the $n^{th}$-order moment
of velocity derivative, i.e. $\overline{(\partial_{x}v_{x})^{n}}\propto
Re^{\rho_{n}}$.
This expression has been studied using high-resolution direct numerical
simulations [1] and it was observed that $A_n$ is  only
weakly dependent on $n$. 
Thus, we will here assume $A_n=C=\text{const}$.
Note that this is consistent with the widely used definition
$Re=\hat{R}e_2=v_{rms}L/\nu$ with $C=1$.
Furthermore, as we will show momentarily, scaling exponents depend only on
$\log (C)$ instead of $C$, which makes potential variations even less
critical in the final result. 

As follows from (1),


\begin{equation}
Re\approx \hat{R}e_{n}^{\frac{n}{\rho_{n}+n}}
\end{equation}

\noindent and,  the widely used large-scale 
``Reynolds number''  $Re\equiv
\hat{R}e_{2}=\frac{(\partial_{x}v_{x})_{rms}L^{2}}{\nu}$ is seen
to be simply one
of the many dimensionless coupling constants characterizing the flow.
One can also introduce an infinite number of Reynolds numbers based
on an infinite number of length scales analogous to the traditional
Taylor microscale:

\begin{eqnarray}
\hat{R}_{\lambda,n}= \sqrt{\frac{5L^{4}}{\overline{3{\cal E}\nu}}}\hat{v}(2n,n)=\nonumber \\
\sqrt{\frac{5v_{0}^{3}}{\overline{3L}{\cal E}}}\sqrt{Re}\times \frac{L^{2}}{v_{0}^{2}}\overline{(\partial_{x}v_{x})^{2n}}^{\frac{1}{n}}\approx Re^{\frac{\rho_{2n}}{n}+\frac{1}{2}}
\end{eqnarray}

\noindent  where the  dimensionless combination ${\cal
E}L/v_{0}^{3}\sim O(1)$. 
Multiplying and dividing the right-hand-side of (3) by viscosity $\nu$  and
taking
into account that in the system under consideration $\overline{{\cal
E}}={\cal P}=const$ (see below),  one obtains
$Re^{\frac{\rho_{2n}}{n}}\approx Re^{\frac{d_{n}}{n}+1}$ giving
$\rho_{2n}=d_{n}+n$.

 \noindent  
It has been demonstrated both theoretically and numerically in Ref.~[1]
that the transitional Reynolds  number  $Re_{2}^{tr}\equiv Re^{tr}$
from ``normal'' to ``anomalous''   scaling of normalized velocity
derivatives 
$\overline{(\partial_{x}v_{x})^{2n}}\propto
(\frac{v_{o}}{L})^{2n}Re^{\rho_{2n}}$  and those  of dissipation rates
$\overline{{\cal E}^{n}}\propto \overline{{\cal E}}^n Re^{d_{n}}$
are  $n$-dependent,  monotonically decreasing with increase of the moment
order $n$ [1] (also see Fig.1).  It was further shown that
$Re^{tr}\approx 100$ ($R^{tr}_{\lambda}\approx 9.$) for $2n\approx 4$.
{However, expressed in terms of $\hat{R}_{\lambda,n}$, the observed
transition points   $\hat{R}^{tr}_{\lambda,n}$ were  $n$-independent with 
  $\hat{R}^{tr}_{\lambda,n}\approx 9.0$.  In this paper, using this
  effect 
   as a dynamic constraint, we generalize the theory developed in [1] to
   calculate anomalous scaling exponents $\rho_{n}$, $\zeta_{n}$  and
   $d_{n}$.}\\

We close this section with Table I which summarizes, 
for convenience, the Reynolds numbers used here 
with their definitions and a description.

\begin{table}[h]
\begin{tabular}{l|p{0.7\textwidth}}
\hline
Reynolds number & Description \\ \hline \hline
$Re =v_{rms} L/\nu$ & Large-scale Reynolds number\\
 $R_\lambda = \sqrt{5/(3{\cal E}\nu)}u_{rms}^2$ & Taylor Reynolds number \\
$Re_{n}^{tr}$ & Large-scale Reynolds number at the transition point 
                     for moments of order $n$ \\
$R_{\lambda,n}^{tr}$ & Taylor Reynolds number at the transition point 
                     for moments of order $n$ \\
$\hat{R}e_n=\hat{v}(n,n)L/\nu$ & Order-dependent Reynolds number; 
probes regions with different amplitudes of velocity gradients\\
$\hat{R}_{\lambda,n}=(5L^4/3{\cal E} \nu)^{1/2}\hat{v}(2n,n)$ &  
Analogous to $\hat{R}e_n$ but based on generalized 
Taylor length scales.
\\
\hline
\end{tabular}
\label{tab:res}
\caption{Summary of Reynolds numbers used in this work.}
\end{table}

 \noindent  {\bf  III. The model}.  Fluid flows can be described by the Navier-Stokes equations subject to boundary and initial conditions (the density is taken $\rho=1$ without loss of generality): 

\begin{equation}\partial_{t}{\bf u}+{\bf u\cdot\nabla u}=-\nabla p +\nu\nabla^{2}{\bf u} + {\bf f} \end{equation}

\noindent   with $\nabla\cdot {\bf u}=0$.   A random Gaussian noise ${\bf f} $ is  defined by the correlation function [7,[8]:

 \begin{equation}\overline{f_{i}({\bf  k},\omega)f_{j}({\bf k'},\omega')}= (2\pi)^{d+1}D_{0}(k)P_{ij}({\bf k})\delta({\hat{k}+\hat{k}'})\end{equation}

\noindent  where the four-vector $\hat{k}=({\bf k},\omega )$ and
projection operator is $P_{ij}({\bf
k})=\delta_{ij}-\frac{k_{i}k_{j}}{k^{2}}$. 
For a fluid in equilibrium 
the  thermal fluctuations, responsible for Brownian motion   are
generated by the forcing (5) with $D_{0}(k)= \frac{k_{B}T
\nu}{\rho}k^{2}$ [7],[8]. In channel flows or boundary layers 
with  rough walls, the amplitude $D_{0}$ is of the order of the rms magnitude of the roughness element [2]. 
Here we are interested in the case $D_{0}(k)=const \neq 0$ only in the interval close to    $k\approx 2\pi/L$, discussed by Forster, Nelson and Stephen  [8].
The energy balance, written here for the case of isotropic and homogeneous flow, following from (5)-(6) imposes the energy conservation constraint:
${\cal P}=\overline{{\bf u}\cdot {\bf f}}=\overline{{\cal E}}=\frac{\nu}{2}\overline{(\frac{\partial u_{i}}{\partial x_{j}}+\frac{\partial u_{j}}{\partial x_{i}})^{2}}= \nu\overline{(\frac{\partial u_{i}}{\partial x_{j}})^{2}}= O(1)$, 
where  the energy production rate  ${\cal P}=O(1)$  is an external
parameter independent of the Reynolds number.   
The random-force-driven NS equation can be written in Fourier space:
\begin{widetext}
\begin{equation}
 u_{l}({\bf k},\omega)=G^{0}f_{l}({\bf k,\omega})-
\frac{i}{2}G^{0}{\cal P}_{lmn}\int u_{m}({\bf q},\Omega)u_{n}({\bf k-q},\omega-\Omega)d{\bf Q}d\Omega
\end{equation}
\end{widetext}
\noindent where $G^{0}=(-i\omega+\nu k^{2})^{-1}$,  ${\cal P}_{lmn}({\bf k})=k_{n}P_{lm}({\bf k})+k_{m}P_{ln}({\bf k})$  and,  introducing the zero-order solution ${\bf u}_{0}=G^{0}{\bf f} \propto \sqrt{D_{0}}$, so that ${\bf u}=G^{0}{\bf f}+{\bf v}$,  one derives the equation for perturbation ${\bf v}$:

\begin{widetext}
\begin{eqnarray}
v_{l}(\hat{k})=-\frac{i}{2}G^{0}(\hat{k}){\cal P}_{lmn}({\bf k})\int v_{m}(\hat{q})v_{n}(\hat{k}-\hat{q})d\hat{q}\nonumber\\
-\frac{i}{2}G^{0}(\hat{k}){\cal P}_{lmn}({\bf k})\int  [v_{m}(\hat{q})G^{0}(\hat{k}-\hat{q})f_{n}(\hat{k}-\hat{q})+ G^{0}(\hat{q})f_{m}(\hat{q})v_{n}(\hat{k}-\hat{q})]d\hat{q} \nonumber \\
-\frac{i}{2}G^{0}(\hat{k}){\cal P}_{lmn}({\bf k})\int G^{0}(\hat{q})f_{m}(\hat{q})G^{0}(\hat{k}-\hat{q})f_{n}(\hat{k}-\hat{q})d\hat{q}
\end{eqnarray}
\end{widetext}

\noindent  It is clear from (7) that the correction to the $O(\sqrt{D_{0}})$ zero-order Gaussian solution is driven 
by  the ``effective forcing''

\begin{eqnarray}
F_{l,2}=-\frac{i}{2}G^{0}(\hat{k}){\cal P}_{lmn}({\bf k}) \times\nonumber\\ \int G^{0}(\hat{q})f_{m}(\hat{q})G^{0}(\hat{k}-\hat{q})f_{n}(\hat{k}-\hat{q})d\hat{q}=O(D_{0})\nonumber
\end{eqnarray}

\noindent which is small in the limit $D_{0}\rightarrow 0$.   We can define $v_{\l}=v_{1,l}+G^{0}F_{l,2}$ etc and generate renormalized expansions in powers of the dimensionless  ``coupling constant'' $\lambda\propto D_{0}$ resembling that formulated in a classic work by  Wyld (see Ref. [9]).  However,  this expansion is   haunted by divergences [5]  and violations of Galileo invariance 
making its analysis  extremely hard if not impossible.\\  

In this work, we avoid perturbative treatments of (6)-(7) and construct 
$M_{2n}= \overline{{\cal E}^{n}}/\overline{{\cal E}}^{n}$
based on   their asymptotic  properties:  
{in the ``weak turbulence'' limit $Re\ll Re_{2n}^{tr}$ 
(where $Re_{2n}^{tr}$ is the 
transitional Reynolds number observed for $2n$-th order
moment of velocity gradients) the Gaussian
solution  $M_{2n}=(2n - 1)!!$ follows directly  from (7). In the
opposite limit   $Re\gg Re_{2n}^{tr}$ we seek for   the moments as:
$M_{2n}\approx A_{2n}Re^{\rho_{2n}}$ with not yet known
amplitudes   $A_{2n}$ and exponents $\rho_{2n}$. The two curves cross at
a transitional Reynolds numbers  $Re_{n}^{tr}$ investigated in detail in
Refs.~[1,10]. Thus,}

\begin{eqnarray}
(2n-1)!!\approx A_{2n}(Re_{2n}^{tr})^{\rho_{2n}}
\end{eqnarray}

\noindent   In Fig.~1,  equations 
 (8) are tested by  direct numerical simulations of the moments of
 dissipation rate $\overline{{\cal E}^{n}}$ vs Reynolds number based on the
 Taylor scale.  We can  see the horizontal lines corresponding to the
 Re-independent normalized Gaussian 
moments $M_{2n}=(2n-1)!! $ for  $2 \leq n \leq 6$.   At
$R_{\lambda}>R_{\lambda,n}^{tr}$,  the assumed solutions
$M_{n}=A_{n}Re^{d_{n}}$,
 with unknown exponents $d_{n}$,   are clearly seen.  Our goal is to find $R_{\lambda,n}^{tr} $ and the scaling exponents.

\begin{figure}[h]
\includegraphics[height=6.0cm]{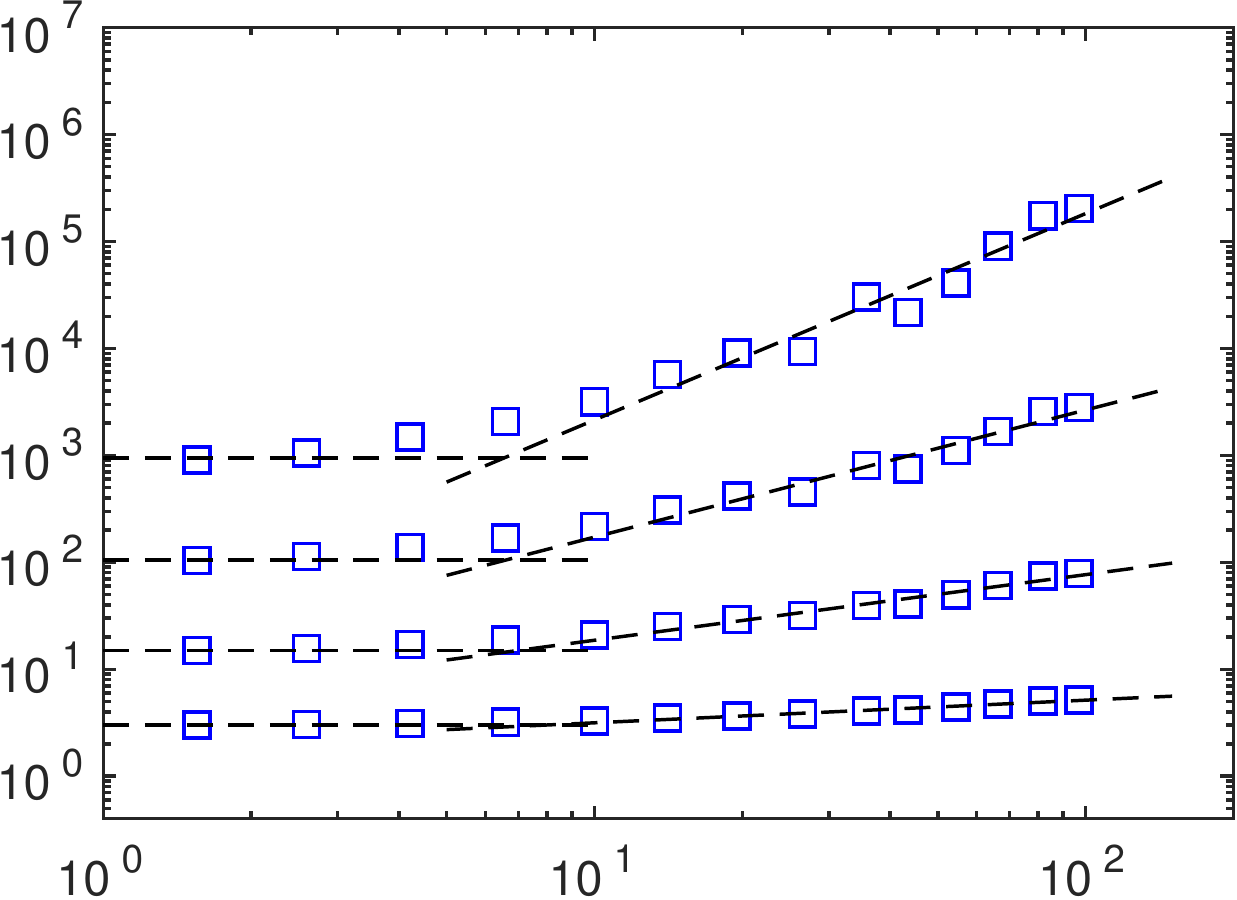}
\begin{picture}(0,0)
\put(-120,-8){$R_\lambda$}
\put(-260,90){$M_{2n}$}
\end{picture}
\caption{ Normalized moments of velocity gradients from direct numerical
solutions to the Navier-Stokes equations (5)-(7). 
From bottom to top $2n=4$, $6$, $8$, and $10$.
Both  asymptotics,
leading to  predicted eq. (8),  are clearly seen. From Ref.~[1].}
\end{figure}
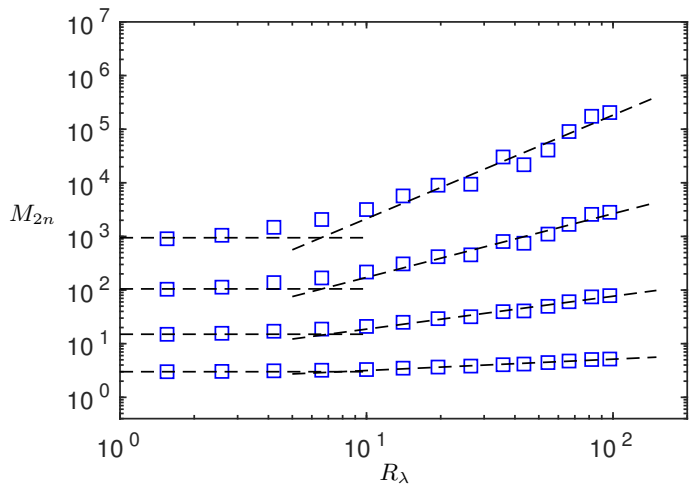

According to  Landau and Lifshitz [7], if $\delta_{\eta}v\approx v(x+\eta)-v(x)$,  the dissipation scale $\eta$ is defined by a condition $Re_{\eta}\approx1=\eta (\delta_{\eta}v )/\nu$, so that $\eta=\nu/(\delta_{\eta}v)$. 
Thus, simple algebra gives: $\partial_{x}v_{x}\approx
(\delta_{\eta}v_{x})^{2}/\nu$. We then obtain
$$(\frac{L}{v_{0}})^{2n}\overline{(\frac{\partial v_{x}}{\partial x})^{2n}}=Re^{2n}(\frac{\eta_{4n}}{L})^{\zeta_{4n}}=Re^{\rho_{2n}}$$
where $Re=v_{0}L/\nu$ and  $\eta_{4n}$ is the viscous cut-off
for velocity structure functions of order $4n$. 
Without loss of generality, we now set the large-scale properties
$v_{0}\equiv v_{rms}=1$ and $L=1$, so that: 

$$
\frac{L^{n}}{v_{0}^{3n}}\overline{{\cal E}^{n}}=Re^{d_{n}}=Re^{n}
\frac{\overline{(\delta_{\eta}v)^{4n}}}{v_{0}^{4n}}
=Re^{n}S_{4n}(\eta_{4n})=Re^{n}(\frac{\eta_{4n}}{L})^{\zeta_{4n}}
$$

\noindent  where the scaling exponents $d_{n}$ and $\rho_{n}$ are yet to be derived from an {\it  a priori} theory 
as well as the exponents $\zeta_{n}$  entering  the so called ``structure functions''  as: $S_{n}(r)=\overline{(v(x+r)-v(x))^{n}}\propto r^{\zeta_{n}}$.   The dissipation rate ${\cal E}=2\nu S_{pq}^{2}$ includes various   derivatives 
$\partial_{p}v_{q}$ with  $p=q$  and $p\neq q$.  Below, based on isotropy, we will use $\overline{{\cal E}}=
15\nu\overline{(\partial_{x}v_{x})^{2}}=v_{0}^{3}/L$  as a normalization
factor yielding  a dimensionless dissipation rate 
$\overline{{\cal E}}v_0/L^3=1$.
This way we assume that all  contributions  to the moments
$\overline{{\cal E}^{n}}$  scale the same way.
Thus, in what follows we
will be working in the units defined by the large-scale properties of
the flow $v_{0}=L=1$ with $Re=1/\nu$.  
In the vicinity of transition, 
 when the forcing is  supported  in a  narrow interval in  the wave-vector space,  $(\partial_{x}v_{x})_{rms}\approx (v_{x}(x+L)-v_{x}(x))_{rms}/L\approx v_{rms}=v_{0}$. In a Gaussian case, where $\rho_{2n}=\rho_{2}n$, all Reynolds numbers are of the same order and 
each one, for example $Re_{1}$,  is sufficient to description the flow
 completely. \\


\noindent {\bf  IV. Transition.}   According to theory and numerical
simulations of Refs.~[1,10,11-14] , $Re^{tr}_{2}\approx 100.0$
or $R^{tr}_{\lambda,2} \approx 9.0$ and  
we associate this Reynolds number  with transition to strong  turbulence,
characterized by non-Gaussian statistics of velocity field and by
anomalous scaling or ``intermittency'' of increments and velocity
derivatives.  
The transition points for high-order moments with $n>2$, expressed in
terms of the standard second-order Reynolds number $Re$,
are observed to satisfy  $Re_n^{tr}<100.0$  ($R^{tr}_{\lambda,n}<9.0$).
 In  turbulent flows the fluctuations with $\hat{v}(n)>v_{rms}$
 (where for simplicity in notation $\hat{v}(n)\equiv\hat{v}(n,n)$) do
 exist and one can expect   transitions when the {\it local}
 $\hat{Re}_{n}=\hat{v}(n)L/\nu \geq 100.0$ even in  
 low Reynolds number subcritical flows  with
 $Re< 100.0$  or $R_{\lambda} <9.0$.  This effect has been 
 supported by numerical
 simulations of isotropic turbulence [1]  (also, see Fig.~1) )  and  in
 experiments  in  ``noisy''  channel flow with randomly rough walls [2].
  
To summarize: critical  Reynolds  numbers
$Re^{tr}=(\partial_{x}v_{x})_{rms}L^{2}/\nu$ 
 for the   $n^{th}$-order  moments  of velocity increments  (spatial
 derivatives)   are  $n$-dependent.  However, when expressed in
 terms of the conditional Reynolds numbers 
$\hat{R}e_{n}$,  based on the characteristic velocities  $\hat{v}(n)$,
the
transition occurs at $\hat{R}e_{n}^{tr}\approx 100$ or
$\hat{R}_{\lambda,n}\approx 9.0$, independent on the moment order $n$.
Thus, we  have found   a  new invariant [1], namely,
a common transition point for moments of any order: 

$$\hat{R}^{tr}_{\lambda,n}= R^{tr}_{\lambda,2}\approx 9.0$$

\noindent  where $\hat{R}_{\lambda,n}$,  is given by expression (3).     


\noindent  2. {\it Evaluation of $R_{\lambda}\equiv R_{\lambda,2}$.} In the linear (Gaussian) regime  only the modes ${\bf v(k)}$ with $k \approx 2\pi/L$ are excited and   in the vicinity of a transition point we can define the Taylor-scale Reynolds number:
 
\begin{eqnarray}
R_{\lambda}\equiv R_{\lambda,2}=\sqrt{\frac{5}{3{\cal E}\nu}}v_{rms}^{2}=
\frac{v_{rms}L}{\nu}\sqrt{\frac{5v_{rms}^{2}\nu}{3{\cal P} L^{2}}}\nonumber \\
 \approx Re\times \sqrt{\frac{5v_{rms}^{3}}{3 {\cal P}LRe}}\approx  \sqrt{Re/1.2}
\end{eqnarray}

 The formulation (5)-(7)  has one important advantage:  statistical properties of the low-Reynolds number flow  (zero-order solution) are externally prescribed by the choice of a random driving force. This means that  it enables one to study transformations  of  different random flows in a controlled way.  Here we restrict ourselves by considering a Gaussian case. 
In  the  low Reynolds number regime  (below transition),  when
$R_{\lambda}<R^{tr}_{\lambda}$,
the integral ($L$), dissipation ($\eta$)  and Taylor ($\lambda$) length
scales are of the same order.    
  
\noindent  Thus,  at a transition point,  where the  theoretically
predicted and supported by numerical simulations
$R_{\lambda,tr}\approx 9$ [1,10,12-14] (also see Fig.1)
we obtain 
$Re^{tr}\approx 1.2 R^{2}_{\lambda,tr}\approx 120$, close to the one
obtained in numerical simulations  [10]. This estimate is based on 
(5)-(6) with a constant, 
Reynolds-number-independent 
dissipation rate $\overline{{\cal E}}={\cal P}=const$ and
a Kolmogorov-like estimate  $u_{rms}^{3}\approx {\cal P}L$ with
the Kolmogorov's constant $C_{K}\approx 1.65$  and at the internal scale
$2\pi/L\approx 1$.\\

\noindent {\bf  V. Relation between exponents $\rho_{n}$ and $d_{n}$
when  $\overline{{\cal E}}={\cal P}=const$.}

In the limit $Re\rightarrow 0$, we have:

$$M_{2n}^{<}=\frac{{\overline{(\partial_{x}v_{x}
)^{2n}}}}{\overline{(\partial_{x}v_{x})^{2}}^{n}}=\frac{\overline{{\cal
E}^{n}}}{\overline{{\cal E}}^{n}}
=(2n-1)!! $$

\begin{figure}[h]
\includegraphics[width=8cm]{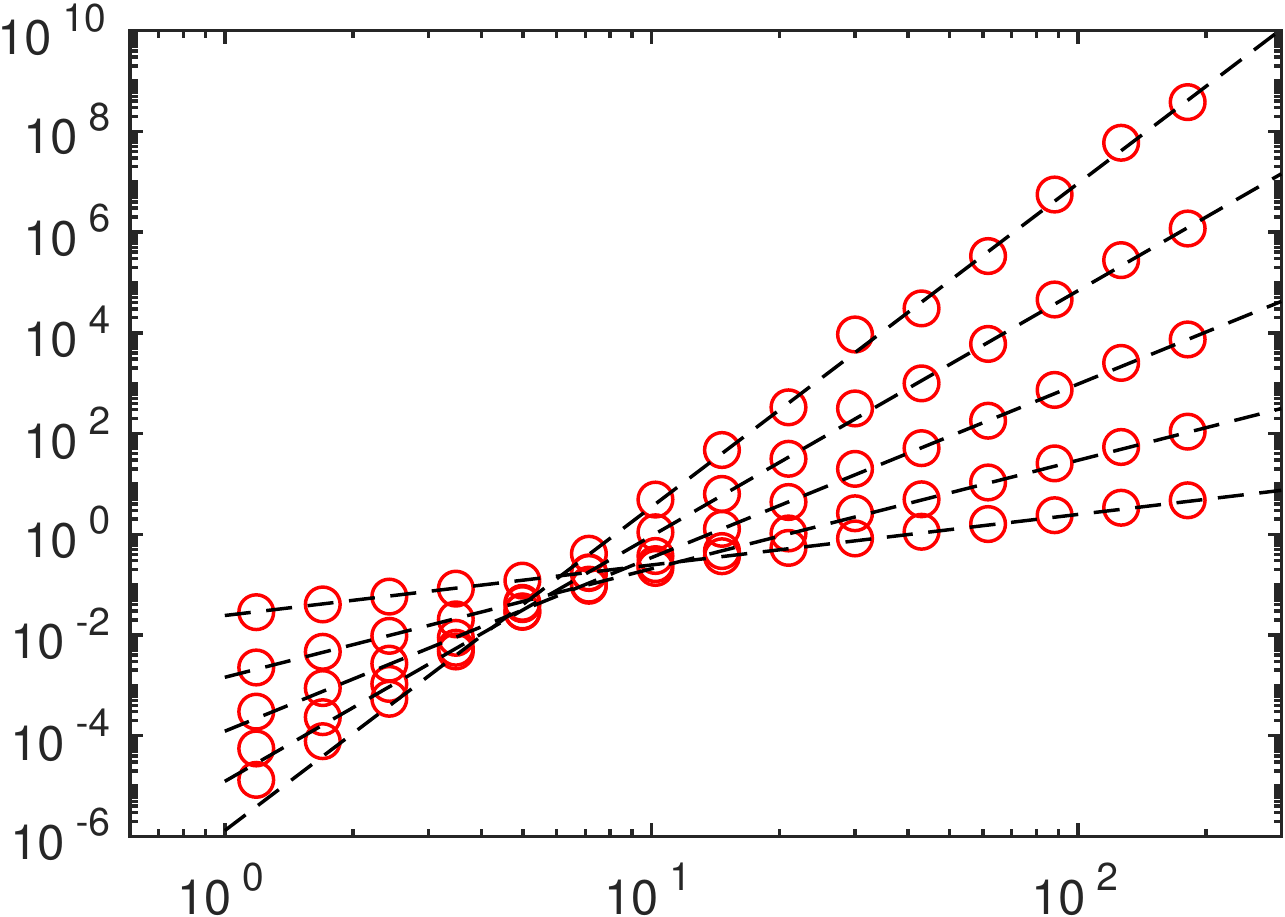}
\begin{picture}(0,0)
\put(-110,-8){$Re$}
\put(-245,60){\rotatebox{90}{$\overline{(\partial_x u_x)^{2n}}(L/v_0)^{2n}$}}
\end{picture}
\caption{ Dimensionless moments of velocity gradients. 
Symbols from simulations in Ref.~[1]. 
Dashed lines are the theoretical predictions.
 }  
 \end{figure}
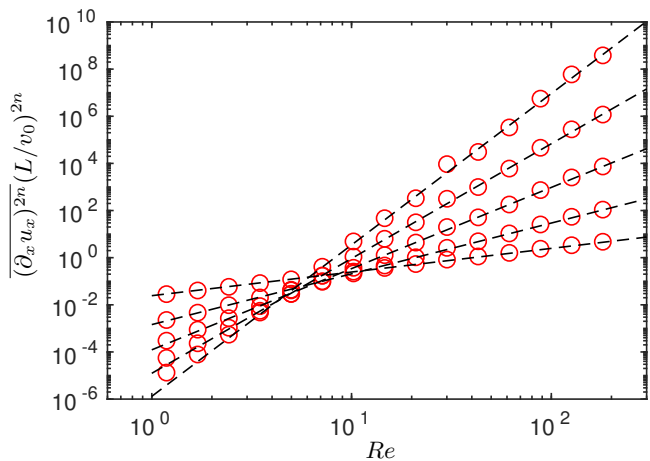

 \noindent  and seek  the large-$Re$ solution of  the form:
 
\begin{equation}
M_{2n}^{>}=\overline{(\partial_{x}v_{x})^{2n}}\propto  (\frac{v_{0}}{L})^{2n}Re^{\rho_{2n}}
\end{equation}

\noindent  
 Multiplying (10) by $\nu^{n}$ gives:
 
 \begin{equation}
 (\frac{L}{v_{0}^{3}})^{n}\overline{{\cal E}^{n}}=Re^{d_{n}}=Re^{\rho_{2n}-n}
 \end{equation}

\noindent leading  to the relation between scaling exponents $d_{n}$ and $\rho_{2n}$:
 
\begin{equation}
\rho_{2n}=d_{n}+n
\end{equation}

\noindent  obtained above. This  relation,   which is is a consequence
of the
energy conservation law in the flow driven by white-in-time random
force, was studied in Ref.~[10].
However, the forcing in [10] was not white-in-time.
To stay closer to the theory, 
here we present results from high-fidelity direct 
numerical simulations (described below), 
with a white-in-time forcing scheme. \\
 


\noindent  {\bf  VI. Scaling exponents.}  
As follows from the definition (3),  at a transition point of the $n^{th}$
moment, the large-scale Reynolds number is:
\begin{equation}
Re^{tr}_{n}=C(\hat{R}^{tr}_{\lambda,n})^\frac{1}{\frac{d_{n}}{n}+\frac{3}{2}}
\end{equation}
\noindent  
Since $R_{1}^{tr}\approx 100-200$, $d_{1}=0$ (see below)  and
$R^{tr}_{\lambda,1}\approx 8.91$,  we obtain  an estimate $C\approx  50-100$. 

The matching condition at the transition Reynolds number, can now be written as
\begin{equation}
\frac{\overline{{\cal E}^{n}}}{\overline{\cal E}^n}=
(2n-1)!!=C^{d_{n}}(\hat{R}^{tr}_{\lambda,n})^{\frac{nd_{n}}{d_{n}+\frac{3n}{2}}}
\end{equation}
which is a closed equation for the anomalous exponents $d_{n}$.   For $n=1$,
the left side  of  relation (14)  is equal to  unity,  giving $d_{1}=0$ and the
normalized dissipation rate $\overline{{\cal E}}=1$, consistent with the
Navier-Stokes dynamics. 

Using large-scale numerical simulations, it  has recently been shown
that, as discussed above, while the large-scale transitional Reynolds number $Re_{n}^{tr}$
depends on the moment order $n$, the one based on a the $\hat{v}(2n,n)$ field,
is independent of $n$ [1] with a value of $\hat{R}^{tr} _{\lambda,n}\approx 9.0$ 
. { This
result can be readily  understood in terms of the dynamics of
transition to turbulence which is not a statistical feature but a
property of each realization 
where $R_{\lambda}>R^{tr}_{\lambda}$. In other words all fluctuations  with
``local'' $\hat{R}_{\lambda,n}\geq Re^{tr}\approx 9.0$ undergo transition to
turbulence. This argument, consistent with Landau's theory of transition,
fixes  the amplitude in relations (14)-(15) and  enables  evaluation of the
scaling exponents by matching two different flow regimes.}
Taking $\ln 8.91\approx 2.19$  gives:

\begin{eqnarray}
d_{n}=-\frac{1}{2}[n(\frac{2.19}{\ln C}+\frac{3}{2})-\frac{\ln (2n-1)!!}{\ln C}]+ \nonumber \\ 
\sqrt{\frac{1}{4} [n(\frac{2.19}{\ln C}+\frac{3}{2})-\frac{\ln (2n-1)!!}{\ln C}]^{2} +\frac{3}{2}n\frac{\ln (2n-1)!! }{\ln C}}
\end{eqnarray}

\noindent where we must use the representation: $(2n-1)!!=\frac{2^{n}}{\sqrt{\pi}}\Gamma(n+\frac{1}{2})$ valid for non-integer values of $n$. 
This expression is plotted in figure 3a for $n\le 6$. 
In part b of the figure we show that 
for  $n \gg 1$, the exponents approach the asymptotic behavior 
$d_{n}\rightarrow 0.19 n\ln n$.\\

\begin{table}
\begin{ruledtabular}
\begin{tabular}{ccccccccccc}
\hline
$d_{n}$  & $T1$ & $DNS$ &  $C90$ & $Exp.$  \\
\hline
$ d_{1} $ & $0.00$ & $0.00$ &  $0.00 $ & $0.00$ \\
$d_{2} $& $0.158 $& $0.149$ & $ 0.187$ & $0.152$\\
$ d_{3}$ &$ 0.49$& $0.443$ & $0.46$& $0.4$ \\
 $d_{4} $&$0.94$&$ 0.89 $& $0.80$ &$ 0.73$ \\
 $d_{5}$ &$1.49$& $1.47$ &$ 1.19$  & $1.1$\\
 \end{tabular}
 \end{ruledtabular}
\caption{Comparison of theoretical predictions for  exponents $d_{n}$,
given by relation (15),  with experimental  data, semi-empirical models
and numerical simulations: $T1$: Theory  [10];  $DNS$;  
present direct numerical
simulations;  $MF$:  multi-fractal theory [10]; 
$C90$: equation (15) with $\ln[C]=4.5$;
{\it Exp.}: $d_{n}$ from (15   with $n!$ instead of $(2n-1)!!$
for the  Low-Re moments.  } 
\end{table}
 
\begin{table}
\begin{ruledtabular}
\begin{tabular}{ccccccccccc}
\hline
 $ \rho_{n} $ & $MF$ &$ C90$& $DNS2$\\
\hline
$\rho_{1} $&  $0.474 $& $0.46$& $0.455$\\
$ \rho_{3}$ &$1.57$& $1.58$ & $1.478$\\
$ \rho_{4}$ & $2.19 $& $2.19$ & $2.05$\\
 $ \rho_{5}$& $2.84$&  $2.82 $& $2.66\pm 0.14$\\
 $ \rho_{7} $& $ 4.20 $& $4.13$ & $3.99\pm 0.65$\\
 \end{tabular}
 \end{ruledtabular}
\caption{Comparison of exponents $\rho_{2n}=d_{n}+n$  with the outcome of numerical  simulations and  semi-empirical models.   $MF$ and $DNS2$:  multi-fractal theory  and numerical simulations [10];  $C90$:   expression (15)  with  the constant $C=90.$ }
\end{table}

\noindent {\bf Evaluation of exponents $\zeta_{n}$}.  
From (15) one obtains  $\rho_{1}=0.46$ and,  using the relation  

$$\rho_{2n}=2n+\frac{\zeta_{4n}}{\zeta_{4n}-\zeta_{4n+1}-1}$$

\noindent with  $\zeta_{3}=1$, (see Ref.[10]),  gives $\zeta_{2}\approx
0.699$ which is different from Kolmogorov's $\zeta_{2}=2/3$. 
We would like to point out that the derivation here 
is not based on geometrical considerations
of e.g.\ the dissipation field, 
and typically removed from the governing equations.
Instead our analysis is based on 
an exact asymptotic state at low $Re$ and a well established asymptotic 
state at high $Re$. In this sense is that we believe
this to be a derivation first of its kind. 
In general, 
\begin{equation}
\nu^{n}\overline{(\partial_{x}v)^{2n}}=\nu^{n}\overline{(\frac{\delta_{\eta}v}{\eta})^{2n}}\approx \nu^{-n}\overline{(\delta_{\eta}v)^{4n}}\propto Re^{n}\eta^{\zeta_{4n}}_{4n}=Re^{n +\frac{\zeta_{4n}}{\zeta_{4n}-\zeta_{4n+1}-1}}\approx Re^{d_{n}}
\end{equation}

\noindent with $d_{n}$ evaluated above.  Thus, we obtain a 
recursion relation:

\begin{equation}
\zeta_{4n+1}=\zeta_{4n}F[n]-1
\end{equation}

\noindent where 

\begin{equation}
F[n]=1-\frac{1}{d_{n}-n}
\end{equation}

\noindent subject to initial condition $\zeta_{2}=0.699$, derived  above.   The solution for $n\leq 14$ is shown on Table III.\\


\noindent {\bf Comparison with direct numerical simulations}.
In Figure 2 and Table II, we compare the exponents obtained 
with new high-fidelity direct numerical simulations  (DNS)
of the Navier-Stokes equations forced at low-wavenumbers
with Gaussian white-noise. The accuracy of the simulations 
have been tested through grid convergence in both space and
time. In order to meaningfully compare with the theory, 
the characteristic length and velocity scales must be independent of 
the Reynolds number. In general, the root-mean-square velocity or the 
integral length do present $Re$-dependences, even if weakly,
in practical simulations.
Thus, we use instead velocity and length scales defined by the 
independent forcing ${\bf f}$ in (4) using its spectrum $E_f(k)$.
We then compute 
the variance of the forcing $\overline{f^2}=\int E_f(k)dk$ and
a length scale $L_f = \int E_f(k)k^{-1} dk/\overline{f^2}$.
Finally a velocity scale can be defined as 
$u_f = \sqrt{3}\overline{f^2}^{1/4}L_f^{1/2}$. We have verified 
these to be indeed independent of Reynolds number and of the 
same order of magnitude as the more traditional integral length scale
and the root-mean-square of the velocity field.
The so normalized gradients are shown in Fig.~2 where we see 
a wide scaling range
where exponents $\rho_{n}$ can be obtained as best fits
(dashed lines). The excellent agreement is seen in Table II.
In Table III we also see good agreement with the popular
multi-fractal formalism and other numerical simulations.

\begin{table}
\begin{widetext}
\begin{ruledtabular}
\begin{tabular}{ccccccccccccccc}
\hline
$n$ & $1$ & $2$ & $3$ & $4$ & $5$ & $6$ & $7$ & $8$ &$ 9$ & $10$ & $11$ & $ 12$ & $13$ & $14$ \\
$\zeta_{n}(T)$ & $0.369$ & $0.699 $ & $1.0$ & $1.26$ & $1.53$ &$1.78$ & $2.07$ & $2.29$  & $2.56$ & $2.84 $ & $3.14$ & $3.47$ & $3.85$ & $4.27$\\
$\zeta_{n}(SA) $& $xx $ & $0.7$ & $ 1.0$ & $1.25$ & $xx$ & $1.8$ & $2.0$ & $2.2 $ & $2.3$ & $2.5$\\
 \end{tabular}
 \end{ruledtabular}
 \caption{Scaling exponents $\zeta_{n}$ of velocity structure functions $S_{n}\propto r^{\zeta_{n}}$.
 (T): Theory developed in this paper; 
 (SA): exponents from Sreenivasan and Antonia~[5]. $xx$ stands for missing  data.
 }
 \end{widetext}
\end{table}

\  \\

\noindent {\bf Kolmogorov's Theory.}  
The Kolmogorov's exponents $\zeta_{n}=n/3$ are obtained from the above theory  for a finite moment order $n$ in the limit $\ln C \rightarrow\infty$.
  As follows from (15),  in this case $d_{n}=0$ 
and the relations (16)-(18)  give  Kolmogorov's exponents  $\zeta_{n}=n/3$.

\begin{figure}[h]
\includegraphics[height=0.27\textwidth]{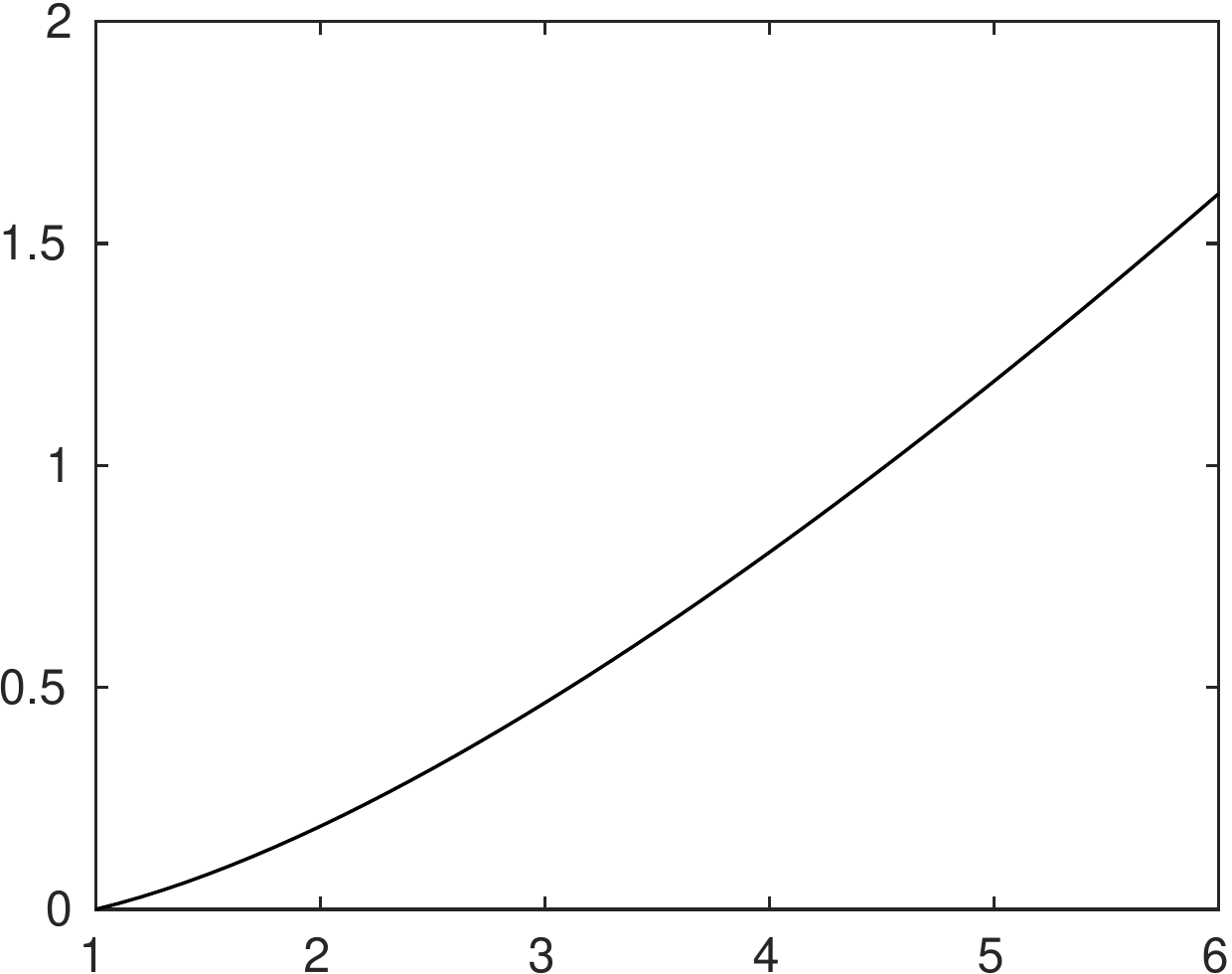} \qquad \qquad
\includegraphics[height=0.28\textwidth]{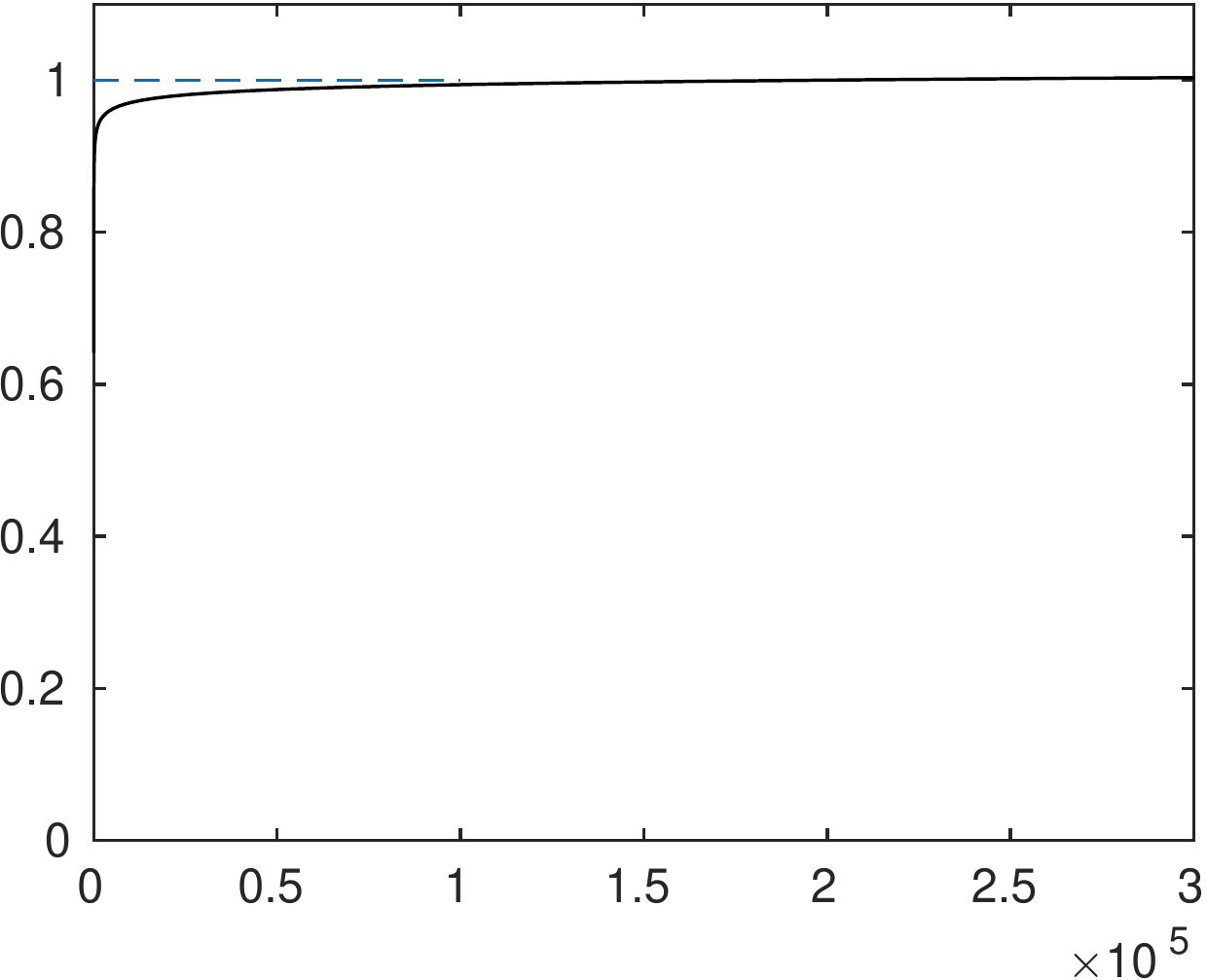}
\begin{picture}(0,0)
\put(-410,80){$d_n$}
\put(-300,-8){$n$}
\put(-90,-8){$n$}
\put(-210,80){${5.2 d_n\over n \ln n}$}
\end{picture}
\caption{  Anomalous exponents $d_{n}$ of the moments of dissipation rate
$\overline{{\cal E}^{n}}/\overline{{\cal E}}^{n}\propto Re^{d_{n}}$ given by
formula (15). (a) $d_n$ for $n\le 6$. (b) High-$n$ limit  
showing ${5.2 d_{n}}/{n\ln n}\rightarrow 1$. }
\end{figure}
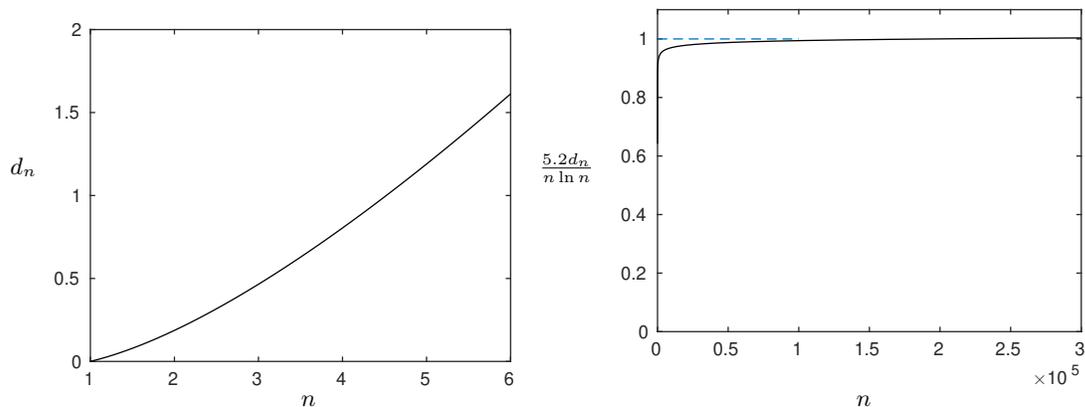

 
 

 
 



\ \\
 
\noindent {\bf  VII. Summary and Conclusions.}   1. This paper is based on
equations (5)-(7) leading to the well-defined Gaussian  low-order moments of
velocity derivatives and increments.    In this case,  transition to strong
turbulence is defined as a first appearance of  anomalous scaling exponents of
the moments of the dissipation rate. No  inertial range enters the
considerations. \\
\noindent 2. In the spirit of Landau's  theory of transition,  {it
is assumed and numerically supported in [1]}  that in each statistical
realization the transition occurs at  $\hat{R}_{\lambda,n}\geq 9$
independent on $n$.    The numerical value of transitional Reynolds
number   $R_{\lambda}\approx 8.908$  was derived in Refs.~[12-14].
Also, this result comes out  of  semi-empirical theories of large-scale
turbulence modeling [14].\\ 
3.  We speculate that  these two   very strong dynamic constraints  are
satisfied by the formation of small-scale coherent structures  {\it manifested
by    intermittency}   (anomalous scaling) of dissipation fluctuations in
turbulence. In addition, the theory yields an energy spectrum  $E(k)\propto
k^{-1.699}$.\\
4.  The universality of these  results is  yet to be studied.  On one hand,
there has been recent support from numerical and experimental data of 
various flows [10,16-17]. On the other hand, as we see from Table.~1,
numerical values of exponents $d_{n}$ may be  sensitive  to  statistics of the
low-Reynolds number fluctuations.  \\  
 One argument in favor of broad universality can be found in Ref.~[8], (Model C),  where dynamic renormalization group was applied to the problem of the Navier-Stokes equations driven by various  random forces. The authors considered a general force (5) of an arbitrary statistics,  supported in a finite interval of wave-numbers $k\approx k_{f}$ and showed  that  in the limit $k<<k_{f}$ the velocity fluctuations, generated by the model,  obeyed Gaussian statistics. This result can be readily understood: each term of the  perturbation expansion of (5)-(7) is $O(k^{2n})$. Therefore, in the limit $k\rightarrow 0$, all high-order contributions with $n>1$ disappear as small. This situation corresponds to weak coupling.  It is not yet clear how universal this result is. \\
5.  The relation (15)  for  anomalous exponents $d_{n}$ is a consequence of  the coupling of  dissipation rate  and the {\bf random } fluctuations of  transitional Reynolds numbers (coupling constant)  studied in Ref.~[1].    The present paper is the first  where the role of  randomness of  a transition point  itself  in  the dynamics of small-scale   velocity fluctuations has been addressed.  It may be of interest to incorporate this feature  in the field-theoretical approaches, like Wyld's diagrammatic  expansions applied to the Navier-Stokes equations for small-scale fluctuations.  \\
6. In this paper  anomalous exponents of moments of velocity derivatives and those of dissipation rate have been calculated without introducing any adjustable parameters. \\
7. It has been  shown in [1] that in the limit of a large moment order $n\gg1 $ the transition to strong turbulence, described in this paper, occurs for $\hat{R}_{\lambda,n}^{tr}\approx 9.0$ and $R_{\lambda}\geq 3$.  The impact of this  constraint  on  numerical values of the exponents $\zeta_{n}$  in the limit $n\rightarrow \infty$ will be investigated in a future communication. \\

 \begin{acknowledgements}     
\noindent V.Y. benefitted a lot from  detailed and illuminating  discussions of this work with  A.M.Polyakov. We are grateful to  H. Chen, G.Eyink,  D. Ruelle, J. Schumacher,  I. Staroselsky, Ya.G. Sinai, K.R. Sreenivasan  and M.Vergassola    for many stimulating and informative discussions.  DD acknowledges support from NSF.
 \end{acknowledgements}          

\begin {references}
 \noindent  1.\  V.~Yakhot and D.A.~Donzis,  ``Emergence of multi-scaling in a random-force stirred fluid'', Phys.\ Rev.\ Lett. {\bf 119},  044501 (2017).\\
\noindent 2.\  C.~Lissandrello,  K.L.~Ekinci and V.~Yakhot, ``Noisy transitional flows in imperfect channels'', J.\ Fluid Mech, {\bf 778}, R3 (2015).\\
 \noindent 3.\ Kuz'min and Patashinskii, Small Scale Chaos at Low Reynolds Numbers'', Preprint 91-20, Novosibirsk, (1991);\ Sov.\ Phys.\ JETP{\bf 49}, 1050 (1979).\\
\noindent  4.\  R.P.~Feynman, ``The Feynman Lectures on Physics'', Addison Wesley Publishing, 1965.  Many sources attribute to Feynman
the following statement: ``... turbulence is  the last unsolved problem of classical physics.'' It was pointed out by G.~Eyink that the original source of this statement  has not been discovered. \\
\noindent  5.\     \ Monin and Yaglom, ``Statistical hydrodynamics'', MIT Press, 1975;  U.\ Frisch, ``Turbulence'', Cambridge University Press, 1995; 
Sreenivasan \& Anotnia, ``The phenomenology of small-scale turbulence'', Ann.\ Rev.\ Fluid Mech.\\ 
\noindent 6.\  A.\ Polyakov, Sov.\ Phys.\ JETP, {\bf 32}, 296 (1971); ``Lectures Given at International  School on High Energy  Physics in Erevan'',
23 November-4 December, (1971);\\
\noindent 7.\  L.D.~Landau and E.M.~Lifshitz, ``Fluid Mechanics'', Pergamon, New York, (1982). \\  
\noindent 8.\ D.~Forster, D.~Nelson and  M.J.\ Stephen,
``Large-distance and long-time properties of a randomly stirred fluid'',
Phys.\ Rev.\ A {\bf 16}, 732 (1977).\\
 \noindent 9.\ H.W.~Wyld, 
 ``Formulation of the theory of turbulence in an incompressible fluid'',
 Ann.\ Phys.\ {\bf 14}, 143 (1961).\\
\noindent 10.\  J.~Schumacher, K.R.~Sreenivasan and V. Yakhot, 
``Asymptotic exponents from low-{R}eynolds-number flows'',
New J.\ of Phys.\  {\bf 9}, 89 (2007).\\
 \noindent 11.\ D.A.\ Donzis, P.K.\ Yeung and  K.R.\ Sreenivasan, 
 ``Dissipation and enstrophy in isotropic turbulence: scaling and resolution effects in direct numerical simulations'',
Phys.\ Fluids {\bf 20}, 045108 (2008).\\ 
 \noindent 12.\ V.\ Yakhot and L.\ Smith, 
 ``The renormalization group, the $\epsilon$-expansion and derivation of
 turbulence models'',
 J.\ Sci.\ Comp.\ {\bf 7}, 35 (1992).\\
\noindent 13.\ V.~Yakhot, Phys.\ Rev.\ E, 
``Reynolds number of transition and self-organized criticality of strong turbulence'',
{\bf 90}, 043019 (2014).\\ 
\noindent 14.\  V.~Yakhot, S.A.\ Orszag, T.\ Gatski, S.\ Thangam and C.\
Speciale, 
``Development of turbulence models for shear flows by a double expansion technique'',
Phys.\ Fluids A{\bf 4}, 1510  (1992);  
B.E.\ Launder and D.B.\ Spalding, ``Mathematical Models of Turbulence'', Academic
Press, New York (1972); 
B.E.\ Launder and D.B.\ Spaulding, 
  ``The numerical computation of turbulent flows'',
  Computer Methods in Applied Mechanics and engineering, {\bf 3}, 269 (1974).\\
\noindent  15.\  P.E.\ Hamlington, D.\ Krasnov, T.\ Boeck and  J.\ Schumacher, 
``Local dissipation scales and energy dissipation-rate moments in channel flow'',
J.\ Fluid.\ Mech.\ {\bf 701}, 419-429 (2012).\\ 
\noindent 16.\  J.\ Schumacher, J.D.\ Scheel, D.\ Krasnov, D.A.\ Donzis, V.\
Yakhot and   K.R.\ Sreenivasan,
``Small-scale universality in fluid turbulence'',
Proc.\ Natl.\ Acad.\ Sci.\ USA {\bf 111}, 10961-10965 (2014).\\
\noindent 17.\ H.~Chen et al., 
``Extended Boltzmann Kinetic Equation for Turbulent Flows'',
Science, {\bf 301} , 633 (2003).

\end{references}

\end{document}